\documentclass[twocolumn,aps,prl,superscriptaddress]{revtex4}
\makeatletter

\newcommand{\Rmnum}[1]{\expandafter\@slowromancap\romannumeral #1@}
\makeatother
\usepackage{amsmath}
\usepackage{graphicx}
\usepackage{subfigure}
\usepackage{color}
\usepackage{amsfonts}
\usepackage{tikz}
\usepackage{esint}
\usepackage{mathrsfs}
\usepackage{hyperref}
\hypersetup{
     colorlinks = true,
     citecolor  = blue,  
     linkcolor  = red 
}
\begin{document}

\title{Novel quantum phases of two-component bosons with pair hopping in synthetic dimension}
\author{Zhi Lin}
\email{zhilin13@fudan.edu.cn}
\affiliation{Department of Physics and State Key Laboratory of Surface Physics, Fudan University, Shanghai 200433, P.R. China}
\affiliation{School of Physics and Materials Science, Anhui University, Hefei 230601, P. R. China}

\author{Chenrong Liu}
\affiliation{Department of Physics and State Key Laboratory of Surface Physics, Fudan University, Shanghai 200433, P.R. China}

\author{Yan Chen}
\email{yanchen99@fudan.edu.cn}
\affiliation{Department of Physics and State Key Laboratory of Surface Physics, Fudan University, Shanghai 200433, P.R. China}
\begin{abstract}
We study two-component (or pseudospin-1/2) bosons with pair hopping interactions in synthetic dimension, for which a feasible experimental scheme on a square optical lattice is also presented. Previous studies have shown that two-component bosons with on-site interspecies interaction can only generate nontrivial interspecies paired superfluid (super-counter-fluidity or pair-superfluid) state. In contrast, apart from interspecies paired superfluid, we reveal two new phases by considering this additional pair hopping interaction. These novel phases are intraspecies paired superfluid (molecular superfluid) and an exotic non-integer Mott insulator which shows a non-integer
atom number at each site for each species, but an integer for total atom number.
\end{abstract}
\maketitle

Ultracold quantum gases are highly controllable systems, in which various novel interaction and detection techniques can be realized, and the extreme physical parameter regimes can be reached \cite{Lewenstein1, Bloch1, AGF, Bloch2, AGF1, Eckardt_eff}. Thus, ultracold quantum gas systems have been used to simulate quantum many-body systems and provide an ideal platform to discover novel quantum states. In bosonic systems, there are two kinds of boson pair condensation states with either the intraspecies pairing \cite{pair_condensation1} or the interspecies pairing \cite{Spin1}. The interspecies paired superfluid state has been proposed in a two-component Bose-Hubbard model with on-site interspecies interaction \cite{Spin1, Demler, Spin2, PSF4, PSF5, PSF6, PSF7, PSF8, PSF9, PSF10}. Moreover, the intraspecies paired superfluid or molecular superfluid (MSF) has also been predicted in three different single-component bosonic systems, i.e., an atomic Bose gas with a Feshbach Resonance \cite{Single-PSF1, Single-PSF2, Single-PSF-lifetime},  attractive Bose-Hubbard model with three-body on-site constraint \cite{Single-PSF3, Single-PSF4} and extended Bose-Hubbard model(EBHM) with pair hopping \cite{pair_hopping1,pair_hopping2,pair_hopping3,pair_hopping4}.

Unfortunately, the MSF in single-component bosonic systems has not been observed experimentally. One reason is the short lifetime of molecular condensates by using the Feshbach resonance technique \cite{Single-PSF-lifetime}. Besides, it is quite difficult to realize the attractive Bose-Hubbard model with a three-body constraint. Moreover, MSF is predicted in EBHM (when $V\!\neq \!0$) under large value of pair hopping $P$ and nearest-neighbor interaction $V$ \cite{pair_hopping1,pair_hopping2,pair_hopping3,pair_hopping4}, but it is hard to reach this parameter region in experiment. In a real experimental system, $P$ and $V$ are much smaller than normal hopping and on-site interaction by 3-4 orders of magnitude \cite{non_standard}. Indeed, the calculation in EBHM ignores the effect of an important term, i.e., density-induced tunneling $T$, which could be much larger than $V$ and $P$. Thus, alternative feasible experimental schemes such as implementing a feasible scheme in the interacting two-component bosonic systems,  are imperiously needed to observe this fascinating MSF state. Meanwhile, there is still a lack of a study on the exotic Mott insulator (MI) phase in the interacting two-component bosons. On the whole, two-component bosons with novel interaction may provide an opportunity for discovering the novel phases.

\begin{figure}[h]
\centering
\includegraphics[width=0.8\linewidth]{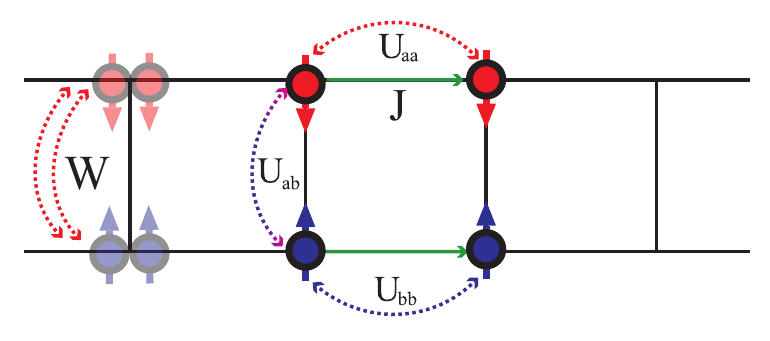}
\caption {This two-component boson system in 1D chain can be mapping to a coupled two-spin chain with SPH interaction $W$. The green arrow indicate intraspecies normal hopping $J$, double both sides dashed arrow indicates SPH interaction $W$,  on-site interaction ($U_{\rm{aa}}$, $U_{\rm{ab}}$, $U_{\rm{bb}}$ ) are indicated by both sides dashed arrow.}
\label{cha3_onsite_pair_hopping}
\end{figure}

On the other hand, by periodically shaking optical lattice \cite{floquet2, Eckardt_eff, Bukov, Goldman,mei} or modulating interaction strength \cite{tunable-so,induce}, Floquet technique has shown its ability to engineer the form and intensity of interactions in various experiments.  So far, Floquet engineering is mainly focused on manipulating the `single-particle hopping' processes \cite{s-hopping1,s-hopping2,s-hopping3,s-hopping4,s-hopping5,s-hopping6,s-hopping7,s-hopping8,s-hopping9,
s-hopping10,s-hopping11,s-hopping12,s-hopping13}, where the hopping amplitude or hopping phase (Peierls phase) depends on the occupation numbers of the sites relevant to hopping processes. The internal atomic degrees of freedom, e.g., pseudospin, can be considered as the synthetic ``dimensions'' \cite{Synthetic_D}. By coupling to a periodically modulating radio-frequency field, a new type of two-particle hopping process with pair hopping interaction along a synthetic dimension or synthetic pair hopping (SPH) interaction (see Fig.~\ref{cha3_onsite_pair_hopping}) can be realized in a two-component boson system.

In this letter, we propose a Floquet engineering scheme in two-component boson system to generate such a new two-particle hopping process with SPH interaction. Two novel quantum states of matter may emerge, including the molecular superfluid (MSF) state and the non-integer Mott insulator (NMI) state. The NMI state displays that the number of the total atoms of two-component at each site is an integer, but each-component is non-integer. This NMI phase may provide a possible platform to discover the exotic magnetic phase. Furthermore, the detection of these two novel states has been addressed. The realization of our scheme provides a basis for further exploration of the exciting many-body phases in synthetic dimensions.

\begin{figure}[h!]
\centering
\includegraphics[width=0.75\linewidth]{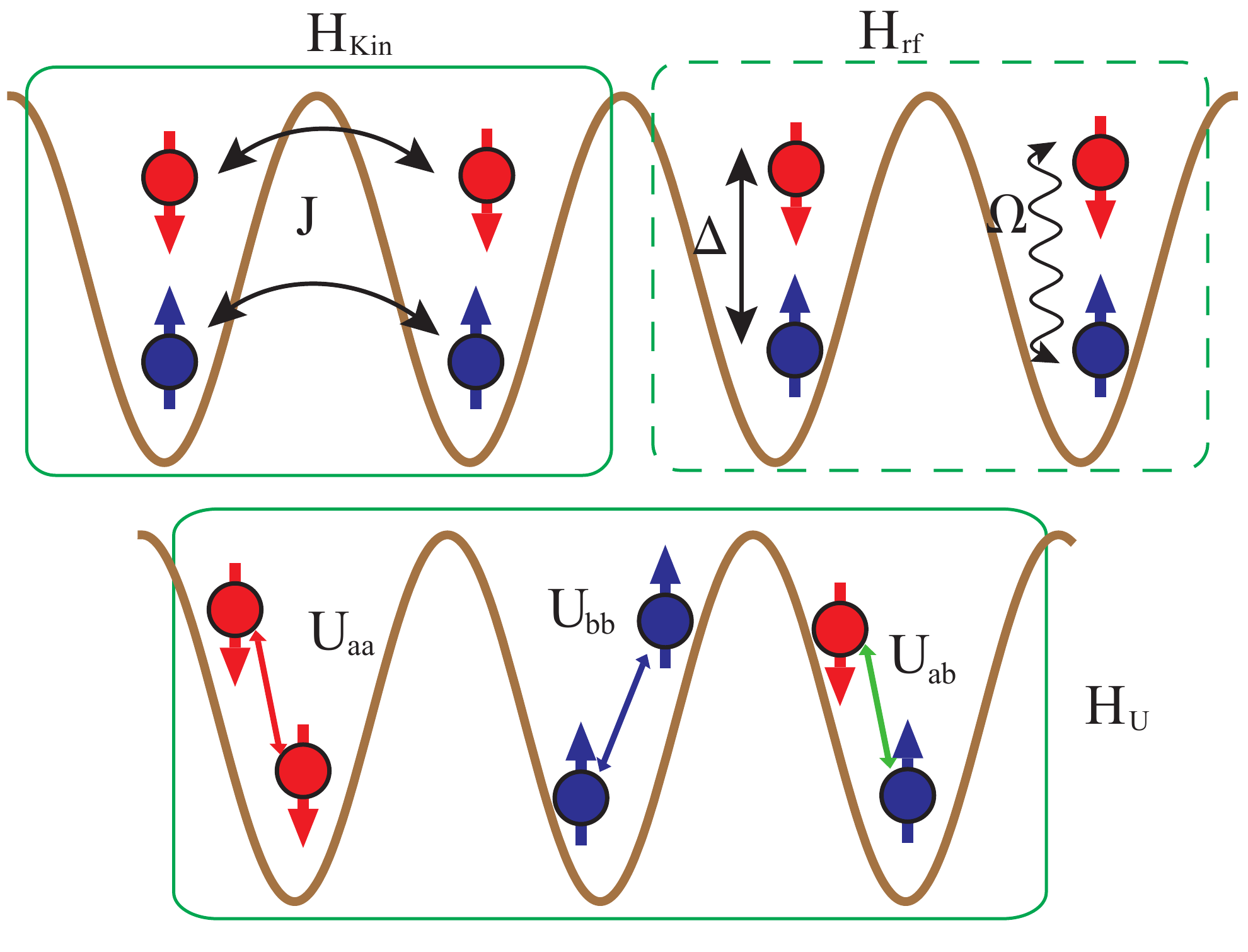}
\caption {The relevant physical processes of time-dependent systems. $\hat{H}_{\rm{Kin}}$ describes each spin hopping between the  nearest neighbour site, $\hat{H}_{\rm{rf}}(t)$  is relevant to radio-frequency coupling of the two spin states with periodic Rabi frequency $\Omega(t)$, and $\hat{H}_{\rm{U}}$ represents the on-site interaction ($U_{\rm{aa}}$, $U_{\rm{bb}}$, and $U_{\rm{ab}}$ labels the strength).  }
\label{schematic_bose}
\end{figure}

\emph{The effective Hamiltonian.---}
We now turn to the realization of  SPH interaction for two-component bosons on square optical lattice, by using periodic modulating radio-frequency field. We firstly introduce the time-dependent Hamiltonian which is used to describe the physics of this periodic modulated two-component boson system. In order to illustrated conveniently and vividly,  the relevant physical processes of this time-dependent systems have shown in one-dimensional (1D) systems (see Fig.~\ref{schematic_bose}). Then the corresponding time-dependent Hamiltonian reads $\hat{H}(t)=\hat{H}_{\rm{Kin}}+\hat{H}_{\rm{rf}}(t)+\hat{H}_{\rm{U}}$, where the on-site interaction contains three terms $\hat{H}_{\rm{U}}=\hat{H}_{\rm{U}}^{\rm{aa}}+\hat{H}_{\rm{U}}^{\rm{bb}}+\hat{H}_{\rm{U}}^{\rm{ab}}$.
Here $\hat{H}_{\rm{Kin}}$ describes normal hopping terms  between nearest neighbour site for each spin and chemical potential, which have the form $\hat{H}_{\rm{Kin}}=-J\sum_{s}(\hat{A}^{\dag}_{s}\hat{A}_{s-1}+H.C.)-\mu\sum_{s}\hat{A}^{\dag}_{s}\hat{A}_{s}$,
where $J$ is the spin-independent hopping amplitude, $\mu$ is spin-independent chemical potential and  $\hat{A}_{s}=(\hat{a}_{s}, \hat{b}_{s})^{T}$  are vector field with annihilation operators $\hat{a}_{s}$ ($\hat{b}_{s}$) on lattice site $s$ for spin-down (spin-up) component. Two spin states coupled by periodic radio-frequency field and the corresponding hamiltonian reads $
\hat{H}_{\rm{rf}}(t)=\left(\hbar\Delta/2\right) \sum_{s}\hat{A}^{\dag}_{s}\hat{\sigma}_{z}\hat{A}_{s}-\left(\hbar\Omega(t)/2\right)\sum_{s}\hat{A}^{\dag}_{s}\hat{\sigma}_{x}\hat{A}_{s}, \label{driven_rf}
$ where $\Delta =\omega_{\rm{res}}-\omega_{\rm{rf}}$ is the detuning of the radio wave ($\omega_{\rm{rf}}$) from the atomic resonance ($\omega_{\rm{res}}$), $\Omega(t)=\Omega\sin\left(\omega t\right)$ is Rabi frequency,
and $\hat{\sigma}_{x,z}$ are pauli matrices \cite{rf}.
Intraspecies and interspecies on-site interactions are denoted by
$H^{\rm{aa}}_{\rm{U}}\!=\!\left(U_{\rm{aa}}/2\right)\!\sum_{s}\!n_{as}\!\left(\!n_{as}-1\!\right)$, $H^{\rm{bb}}_{\rm{U}}\!=\!\left(U_{\rm{bb}}/2\right)\!\sum_{s}\!n_{bs}\!\left(\!n_{bs}-1\!\right)$,
$H^{\rm{ab}}_{\rm{U}}\!=\!U_{\rm{ab}}\sum_{s}n_{as}n_{bs}$,
where $U_{\rm{aa}}$, $U_{\rm{bb}}$, and $U_{\rm{ab}}$ labels the strength of the on-site repulsive interactions.

Then we obtain the effective Hamiltonian (see a derivation in the Supplemental Material (SM) \cite{supplemental}.)
\begin{eqnarray}
\hat{H}_{\rm{eff}}
\!&\!=\!&\!-\!J\sum_{s}\!\left(\!\hat{a}^{\dag}_{s}\hat{a}_{s-1}+\hat{b}^{\dag}_{s}\hat{b}_{s-1}+H.C.\!\right)\!
-\!\mu\!\sum_{s}\!\left(\!\hat{n}_{as}+\hat{n}_{bs}\!\right)\!\nonumber \\
&&+\frac{U_{\rm{aa}}^{\rm{eff}}}{2}\!\sum_{s}\!\hat{n}_{as}\!\left(\!\hat{n}_{as}\!-\!1\right)\!+
\frac{U_{\rm{bb}}^{\rm{eff}}}{2}\!\sum_{s}\!\hat{n}_{bs}\!\left(\!\hat{n}_{bs}-1\!\right)\! \nonumber \\
&&\!+U_{\rm{ab}}^{\rm{eff}}\!\sum_{s}\!\hat{n}_{as}\hat{n}_{bs}
\!+\!W\!\sum_{s}\!\left(\!\hat{a}^{\dag}_{s}\hat{b}_{s}\hat{a}^{\dag}_{s}\hat{b}_{s}\!+\!b^{\dag}_{s}\hat{a}_{s}\hat{b}^{\dag}_{s}\hat{a}_{s}\!\right)\!, \label{eff-Hamil}
\end{eqnarray}
where the preceding five terms describe two-component Bose-Hubbard model \cite{Spin1,Demler,Spin2} and the $W$ term represents the processes of SPH along a synthetic dimension.
Here, the effective on-site interaction strength and SPH interaction in  Eq.~(\ref{eff-Hamil}) are given by
$\!U_{\rm{aa}}^{\rm{eff}}\!\!=\!U_{\rm{aa}}\!-\!\left[\Omega/(2\omega)\right]\!^{2}\!\left(\!U_{\rm{aa}}\!-\!U_{\rm{ab}}\!\right)$,
$U_{\rm{ab}}^{\rm{eff}}\!\!=\!\!U_{\rm{ab}}\!+\! 2\Delta U\!\left[\Omega/(2\omega)\right]\!^{2}$, $U_{\rm{bb}}^{\rm{eff}}\!\!=\!\!U_{\rm{bb}}\!-\!\left[\Omega/(2\omega)\right]\!^{2}\!\left(\!U_{\rm{bb}}\!-\!U_{\rm{ab}}\!\right)$,
$W=-\left(\Delta U/2\right)\left[\Omega/(2\omega)\right]^{2}$,
$\Delta U=\left(U_{\rm{aa}}+U_{\rm{bb}}\right)/2-U_{\rm{ab}}$.

To reveal the relevant physical processes of this effective Hamiltonian more clearly, we choose a 1D system as an example, where it can be mapped to a coupled two-spin chain (synthetic chain) systems, and every single chain represents one specie of boson. The relevant processes are shown in  Fig.~\ref{cha3_onsite_pair_hopping}. Although this interesting Hamiltonian in Eq.~(\ref{eff-Hamil}) are obtained with detuning $\Delta =0$, we can also obtain it with an effective detuning $\hbar \Delta_{\rm {eff}}=\hbar \Delta -\left(\mu_{a}-\mu_{b}\right)=0$ even if detuning $\Delta \neq 0 $. This condition can be satisfied by tunning $\mu_{a}$ and $\mu_{b}$ via changing fillings $n_{a}$ and $n_{b}$.

\begin{figure}[h!]
\centering
\includegraphics[width=0.8\linewidth]{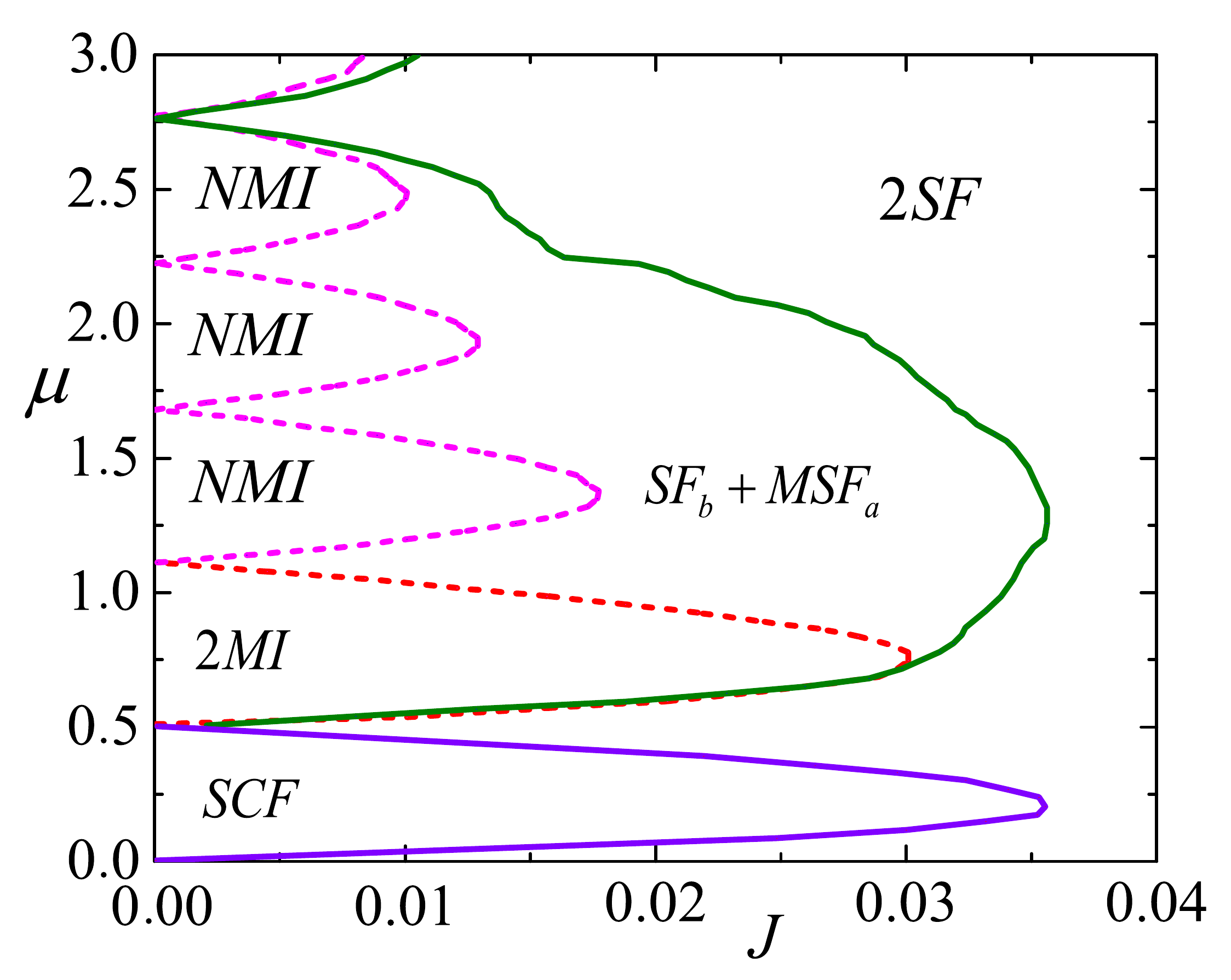}
\caption {The phase diagram of  two species Bose gases with SPH
interation $W$ in square optical lattice. The interaction paraments are $U_{\rm{aa}}=1.0$, $U_{\rm{bb}}=0.7$, $U_{\rm{ab}}=0.5$, and $W=-0.1$.
There are five phases, moveover a novel NMI is new phase have not been researched up to now.}
\label{phase_dia1}
\end{figure}

\emph{The phase diagrams.---}
At below, the phase diagrams will be numerically studied by the Gutzwiller method that has been successfully used to study various phenomena such as stationary states \cite{Gutzwiller1, Gutzwiller2, Gutzwiller3}, time evolution \cite{Gutzwiller4, Gutzwiller5, Gutzwiller6} and excitation dynamics \cite{excitation}. We will use the cluster Gutzwiller method \cite{Gutzwiller7}, which can well capture the quantum fluctuations for a larger cluster to obtain the phase diagrams of the two-component boson gases with SPH interaction on a square optical lattice.
We can naively assume that there exist the nontrivial molecule superfluid state ($\langle\hat{a}_{i}\rangle=0$ but $\phi_{\rm{Da}}=\langle\hat{a}_{i}\hat{a}_{i}\rangle\neq0$) apart from the phases which has been found in the two-component Bose systems with $W=0$. The previous research on the two-component boson system with zero SPH interaction reveals that the asymmetric case ($U_{\rm{aa}}\neq U_{\rm{bb}}$) shows rich phases than the symmetric one ($U_{\rm{aa}}=U_{\rm{bb}}$) \cite{Spin2}. Thus, we study the phase diagrams for the asymmetric case of two-component boson system with finite SPH interaction.
We have chosen a typical asymmetric case $U_{\rm{aa}}=1.0$, $U_{\rm{bb}}=0.7$, $U_{\rm{ab}}=0.5$, and $W=-0.1$ to study the phase diagrams via calculating various possible superfluid orders. The phase diagram is presented in Fig.~\ref{phase_dia1}, where we choose the cluster as $1\times 2$.

There are five phases, i.e., $2$MI, SCF, $2$SF NMI, and SF$_{b}$+MSF$_{a}$ ($\psi_{\rm{a}}\!=\!0$, \!$\psi_{\rm{b}}\!\neq \!0$, \!$\phi_{\rm{Da}}\!\neq\!0$,\! and $\phi_{\rm{Db}}\!\neq\!0$). The $2$SF, $2$MI, and SCF phases have been discussion \cite{Spin1,Spin2}, but NMI phase and SF$_{b}$+MSF$_{a}$ are nontrivial phase which have rarely been predicted in two-component boson systems. Surprisingly, there is no SF$_b$+MI$_{a}$  phase which usually exists in two-component Bose-Hubbard model for the asymmetric case \cite{Spin2}. Transiting from the NMI phase by increasing the value of tunneling amplitude $J$, systems go into an intriguing SF$_b$+MSF$_{a}$ phase which can exist in the lager parameter regions of phase diagrams (see Fig.~\ref{phase_dia1}). In this parameter region, if we switch off the SPH interaction ($W=0$), the SF$_{b}$+MSF$_{a}$ phase will become SF$_b$+MI$_{a}$ phase.  In this sense, the SF$_b$+MI$_{a}$ phase can be considered as the matrix phase of SF$_{b}$+MSF$_{a}$ phase.  In brief, NMI and MSF$_{a}$ phase are induced by the intriguing SPH interaction $W$. At below, we will analyse the property of  NMI and MSF$_{a}$ phases, respectively.

\begin{figure}[h!]
\centering
\includegraphics[width=1.0\linewidth]{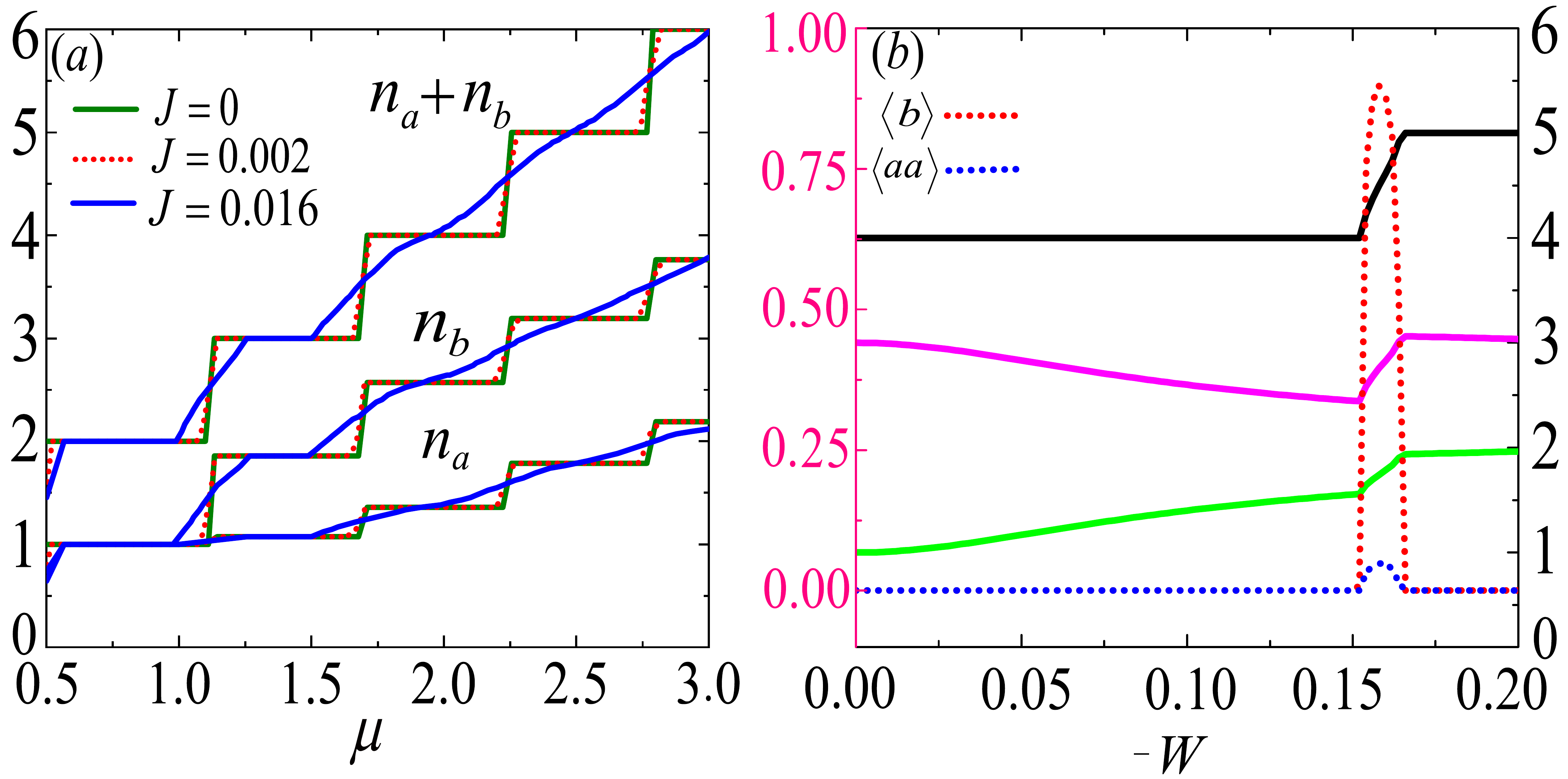}
\caption {(a) The total particle number $n_{a}+n_{b}$, the number of spin-down (spin-up) component $n_{a}$ ($n_{b}$) as a function of chemical potential $\mu$  with $J=0,\, 0.002, \, 0.016$. (b) $n_{a}+n_{b}$, $n_{a}$ and $n_{b}$ as a function of  $W$ with $J=0.002$ and $\mu=2.0$, meanwhile the superfluid order parameter $\langle \hat{a}\hat{a}\rangle$ and $\langle \hat{b}\rangle$ as a function of $W$ are also shown, where the pink (black) vertical axis are indicated the value of superfluid order parameter (particle number). Here the interaction paraments are $U_{\rm{aa}}=1.0$, $U_{\rm{bb}}=0.7$, $U_{\rm{ab}}=0.5$ for figure (a) and (b).}
\label{cha3_particle}
\end{figure}

This nontrivial NMI phase is incompressible, and has nontrivial density distribution feature which shows an integer total atom number at each site while a non-integer atom number for each species. This distribution feature of NMI phase is significantly different from the atom distribution of 2MI phase, and the atom distribution of each site as a function of variation $\mu$  with hopping amplitude $J$ fixed is presented in Fig.~\ref{cha3_particle}(a). The reason why there exist such intriguing NMI phase is that in the limit $J\!=\!0$ (NMI phase), the total number $\hat{n}_{i}$ is a good quantum number but $\hat{n}_{ai}$ and $\hat{n}_{bi}$ are not, since the Hamiltonian $\hat{H}_{J=0}$ commutes with  $\hat{n}_{i}$ but does not commute with $\hat{n}_{bi}$ or $\hat{n}_{ai}$. For the $J\!<\!J_{\rm{critical}}$ case, the property of ground state is unchanged,  but parameter region is shrunken, thus the ground state is also the NMI phase.  Furthermore, the intriguing non-integer feature of NMI phase provides a possible platform to discover a variety of  the interesting magnetic phases.

By changing the value of $W$ and keeping intensity of the other interactions, the systems can evolve from the NMI phase into the novel SF$_{b}$+MSF$_{a}$ phase [see Fig.~\ref{cha3_particle}(b)], where the SF$_{b}$+MSF$_{a}$ phase is characterized by normal superfluid of $b$ (spin-up) component and nontrivial MSF of $a$ (spin-down) component. The characteristics of the MSF$_{a}$ can partly be understood via the coherent state. It's well known that the coherent state satisfies the condition  $\psi_{\rm{a}}\neq 0$ and $\psi_{\rm{Da}}\neq 0$, and even or odd coherent state \cite{even_coherent,odd_coherent} satisfies the condition $\psi_{\rm{a}}=0$ and $\psi_{\rm{Da}}\neq 0$, where even and odd coherent state read $(|0\rangle +\cdots+\alpha^{2n}|2n\rangle/\sqrt{(2n)!}~\!\!)/\cosh|\alpha|^{2}$ and $([\alpha|1\rangle +\cdots+\alpha^{2n+1}|2n+1\rangle /\sqrt{\left(2n+1\right)!}]/\sinh|\alpha|^{2})$, respectively. As is well known, the perfect superfluid phase (the ground state of the Bose-Hubbard model for non-interaction limit $U\!=\!0$) is the coherent state,  but the superfluid phase (in the case of $U \neq 0$)  is not the coherent state \cite{Bloch1}. Thus, the perfect MSF$_a$ can be considered as an odd or even coherent state, but  MSF$_{a}$  state is no longer an even or odd coherent state for interacting systems.

\emph{Symmetry analysis.---} Here we analyse the general symmetry feature of the phases and transitions between them. It is obvious that a finite SPH interaction $W$ breaks $U(1)\times U(1)$ symmetry of the trivial two-component boson Hamiltonian ($W\!=\!0$) down to
 $U(1)\!\times \!Z_{2}$ symmetry (Under the phase transformations $\hat{b}_{i}\rightarrow\hat{b}_{i}e^{i\theta}$ and $\hat{a}_{i}\rightarrow \hat{a}_{i} e^{i\theta}$ [or $\hat{a}_{i}\rightarrow \hat{a}_{i}e^{i(\theta+\pi)}$], the Hamiltonian in Eq.~(\ref{eff-Hamil}) keep unchanged). Here $2$MI and NMI phases break no symmetry, but the SCF, SF$_{b}$+MSF$_{a}$ and $2$SF phases are related to  different ways that the $U(1)\!\times\!Z_{2}$ symmetry is broken. More specifically, the SCF phase breaks discrete $ Z_{2}$  subgroup but the $U(1)$ symmetry is remaining. The SF$_{b}$+MSF$_{a}$ phase breaks $U(1)\!\times \!Z_{2}$ symmetry except for the special point $\theta =\pi$, where SF$_b$ order changes sign ($\langle \hat{b}\rangle \rightarrow \langle\hat{b}\rangle e^{i\pi}$) and MSF$_a$ order keeps unchanged ($\phi_{Da}\rightarrow \phi_{Da}e^{i4\pi}$ and $\phi_{Da}\rightarrow\phi_{Da}e^{i2\pi}$). This type of symmetry breaking is rarely revealed in natural condensed-matter systems. The $2$SF  phase totally breaks the  $U(1)\!\times \! Z_{2}$ symmetry.

\emph{The effective-field analysis of the possible phases.---} We will qualitatively analyze the reason why such rich phases can exist in two-component bosons with SPH interaction. In a $W=0$ case, the mean-field phase diagrams can be obtained by minimizing the free energy $\mathcal{F}_{0}$ of two-component Bose-Hubbard model \cite{Spin2}. The corresponding  phase diagrams can be divided into two typical cases: if the interaction is symmetric, there are three phases, i.e., $2$SF, $2$MI and SCF ($U_{\rm{ab}}\!>\!0$) \cite{Spin1}; if the interaction is asymmetric, the possible phases are $2$SF, $2$MI, SCF and SF$_b$+MI$_{a}$) \cite{Spin2}.  For the $W\neq 0$ case, we can also use the effective field theory to analyze the possible phases of this system.
We can assume the free energy $\mathcal{F}$ has the form (see SM) \cite{supplemental}
\begin{eqnarray}
\mathcal{F}&\!\!=\!\!&\!\mathcal{F}_{0}\!+\!\frac{1}{2}\!\left[\!r_{\rm{Da}}|\phi_{\rm{Da}}|^{2}\!+\!r_{\rm{Db}}|\phi_{\rm{Db}}|^{2}\!+\!
r_{\rm{DD}}\!\left(\!\phi_{\rm{Da}}^{\ast}\phi_{\rm{Db}}\!+\!H.c.\!\right)\!\right]\! \nonumber\\
\!&\!&\!+\frac{1}{4}\!\left[\!g_{\rm{Da}}|\phi_{\rm{Da}}|^{4}\!+\!g_{\rm{Db}}|\phi_{\rm{Db}}|^{4}\right]\!
\!-\!g\!\left(\!\phi_{\rm{SCF}}^{\ast}\psi_{\rm{A}}^{\ast}\psi_{\rm{B}}+H.c.\!\right)\!\nonumber \\
&&-g^{\prime}\!\left(\!\phi_{\rm{Da}}^{\ast}\psi_{\rm{B}}\psi_{\rm{B}}+\phi_{\rm{Db}}^{\ast}\psi_{\rm{A}}\psi_{\rm{A}}+H.c.\!\right)\! \label{eff_free_energy}
\end{eqnarray}
with the condition $ r_{\rm{Da}}>0$, $r_{\rm{Db}}>0$, $g_{\rm{Da}}>0$, $g_{\rm{Db}}>0$.
Here the notation  $\phi_{\rm{Da}}$ ($\phi_{\rm{Db}}$) is  MSF order of the spin-down (spin-up) component. For the asymmetric case ($U_{\rm{aa}}\!>\!U_{\rm{bb}}$), there are four phases, i.e., $2$SF, $2$MI, SCF and SF$_{b}$+MSF$_{a}$ which satisfy the corresponding saddle point equations \cite{supplemental}. Three of them ($2$SF, $2$MI, SCF) have been predicted in a two-component boson system without SPH interaction. Surprisingly, the phase SF$_{b}$+MI$_{a}$ can not exist in this two-component boson system with SPH interaction, and it is replaced by the interesting phase SF$_{b}$+MSF$_{a}$ which has not been predicted in two-component boson system without SPH interaction. This conclusion is in good agreement with numerical calculation. Still, the reason for the existence of NMI can not be revealed by the effective-field analysis (EFS), owing to EFS unable to capture the information of the atom distribution.
\begin{figure}[h!]
\centering
\includegraphics[width=0.8\linewidth]{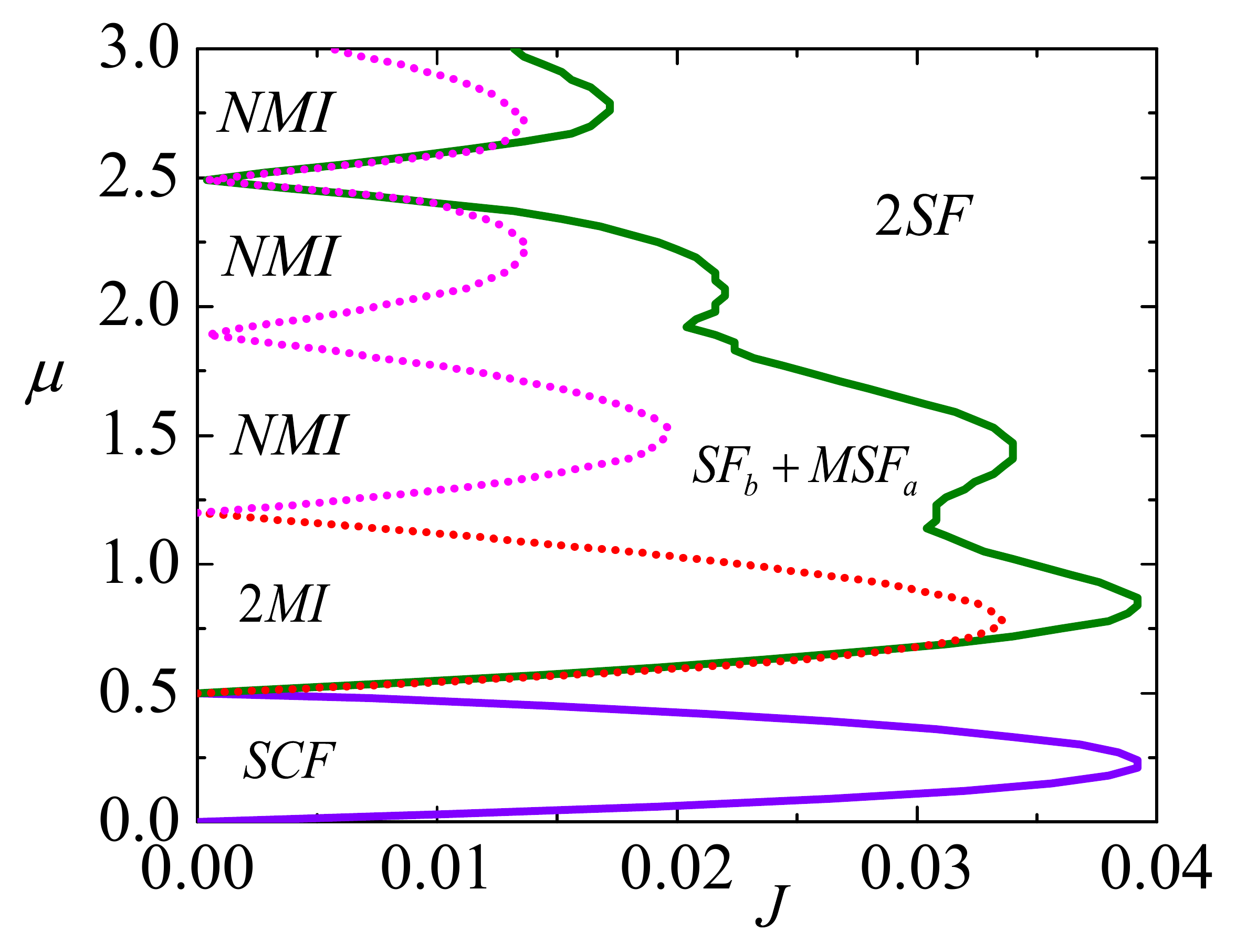}
\caption {The phase diagram of  two species Bose gases with SPH
interation $W$ in square optical lattice. The interaction paraments are $U_{\rm{aa}}=1.0$, $U_{\rm{bb}}=0.7$, $U_{\rm{ab}}=0.5$, and $W=-0.0117$. }
\label{phase_dia1RealW}
\end{figure}

\emph{Experimental realization and detection.---}
If we choose the $\left[\Omega/(2\omega)\right]^{2}\!=\!0.05\!\ll\!1$, only the interesting SPH interaction is important and the high-order terms ($\mathcal{O}\!\left[\!f^{4}(t)/\hbar^{4}\right]$) are ignored (see SM) \cite{supplemental}, then Hamiltonian in Eq.~(\ref{eff-Hamil}) can adequately describe all relevant processes physics of this periodic driving system. If we want to find NMI and MSF$_a$ phase in  a real experimental
system, $W\propto \left[\Omega/(2\omega)\right]^{2}$ must be far less than on-site interaction. By choosing $U_{aa}=2939/2850$, $U_{bb}=2039/2850$, $U_{ab}=61/150$ (can be realized via a Feshbach resonance) and $\left[\Omega/(2\omega)\right]^{2}=0.05$, the effective on-site interactions have the same vales as the vales presented in Fiq.~(\ref{phase_dia1RealW}), while the $W=-0.0117$ is far less than on-site interaction. In this feasible region, NMI and MSF$_a$ phases can also occupy a larger region in the phase diagram, thus the prospects of observing NMI and MSF$_a$ states within this interesting driving system is much larger. Moreover, the nontrivial feature of the number distribution of NMI state can be directly detected by combining spin-removal technique \cite{spin-removal,spin-removal1} and in-situ imaging techniques \cite{In-situ-imaging} which are been successfully to detect the bosonic MI \cite{In-situ-B-MI1,In-situ-B-MI2,In-situ-B-MI3} and fermionic MI \cite{In-situ-F-MI1,In-situ-F-MI2} with single-atom and single-site resolution. The previous research has shown that the MSF and SF phases are distinguished via time-of-flight (TOF) shadow images \cite{Single-PSF-lifetime}, thus SF$_{b}$+MSF$_{a}$ can be directly detected by spin-resolved TOF images \cite{spin-resolved-TOF}.

\emph{Discussion and Conclusions.---}
We have theoretically proposed to engineer a new two-particle hopping process with an SPH interaction in the two-component boson system via periodically modulating the radio-frequency field. This intriguing SPH interaction can lead to two interesting phases, i.e., NMI and MSF$_a$. The NMI state is a new type of Mott insulator, in which the total number at each site is an integer, but each component is a non-integer. The MSF$_a$ state has been proposed for some years, and no much progress has been made to host such a state in a realistic system. The region of NMI and MSF$_a$ states are small shrunken with rapidly decreasing the SPH interaction [see Figs.~(\ref{phase_dia1}, \ref{phase_dia1RealW})]. Thus, the prospects of observing the interesting NMI and MSF$_a$ states are optimistic in a realistic system. Furthermore, the detection schemes of these two novel phases are also addressed. The realization of our scheme provides a possible platform for further exploration of intriguing magnetic phases and interesting many-body phases in synthetic dimensions.

\begin{acknowledgments}
We thank H. Pu, J. B. Gong, T. Qin, and W. L. Liu for helpful discussions. This work is supported by the SKP of China (Grant Nos. 2016YFA0300504 and 2017YFA0304204 ), the NSFC (Grant Nos.11474064, 11625416 and 11947102), the Ph.D. research startup foundation of Anhui University (Grant No. J01003310) and the open project of the State Key Laboratory of Surface Physics at Fudan University (Grant No. KF$2018\_13$).
\end{acknowledgments}


\begin{thebibliography}{99}
\bibitem{Lewenstein1} M. Lewenstein, A. Sanpera, V. Ahufiner, B. Damski, A. Sen, and U. Sen, Adv. Phys. \textbf{56}, 243 (2007).
\bibitem{Bloch1}I. Bloch, J. Dalibard, and W. Zwerger, Rev. Mod. Phys. \textbf{80}, 885 (2008).
\bibitem{AGF} J. Dalibard, F. Gerbier, G. Juzeli\={u}nas, and P. \"{O}hberg,  Rev. Mod. Phys. \textbf{83}, 1523 (2011).
\bibitem{Bloch2} I. Bloch, J. Dalibard, and S. Nascimb\`{e}ne, Nature Physics 8 (4), 267 (2012).
 \bibitem{AGF1}N. Goldman, G. Juzeli\={u}nas, P. \"{O}hberg, and I. B. Spielman,  Rep. Prog. Phys. \textbf{77}, 126401 (2014).
\bibitem{Eckardt_eff}A. Eckardt, Rev. Mod. Phys. \textbf{89}, 011004 (2017).

\bibitem{pair_condensation1}P. Nozi\`{e}res and D. Saint James, J. Phys. (Paris) 43, 1133 (1982).
\bibitem{Spin1}A. B. Kuklov and B.V. Svistunov, Phys. Rev. Lett.\textbf{90}, 100401 (2003).
\bibitem{Demler}E. Altman, W. Hofstetter, E. Demler and M. D. Lukin, New J. Phys. \textbf{5}, 113 (2003).
\bibitem{Spin2}A. Kuklov, N. Prokof\'{e}v, and B. Svistunov, Phys. Rev. Lett.\textbf{92}, 050402 (2004).
\bibitem{PSF4}A. Arg\"{u}elles and L. Santos, Phys. Rev. A \textbf{75}, 053613 (2007).
\bibitem{PSF5} E. K. Dahl, E. Babaev, and A. Sudbo, Phys. Rev. Lett. \textbf{101}, 255301 (2008).
\bibitem{PSF6}S. G. S\"{o}yler, B. C.  Sansone , N. V. Prokof'ev, and B. V. Svistunov, New J. Phys. \textbf{11}, 073036 (2009).
\bibitem{PSF7} A. Hubener, M. Snoek, and W. Hofstetter, Phys. Rev. B \textbf{80}, 245109 (2009).
\bibitem{PSF8} L. Mathey, I. Danshita, and C. W. Clark, Phys. Rev. A \textbf{79}, 011602(R) (2009).
\bibitem{PSF9} A. Hu, L. Mathey,  I.  Danshita, E.  Tiesinga,  C.  J. Williams,  and C. W. Clark, Phys. Rev. A \textbf{80}, 023619 (2009).
\bibitem{PSF10}C. Menotti and S. Stringari Phys. Rev. A \textbf{81}, 045604 (2010).
\bibitem{Single-PSF1}L. Radzihovsky, J. Park, and P. B.Weichman, Phys. Rev. Lett. \textbf{92}, 160402 (2004).
\bibitem{Single-PSF2}M. W. J. Romans, R. A. Duine, Subir Sachdev, and H. T. C. Stoof, Phys. Rev. Lett. \textbf{93}, 020405 (2004).
\bibitem{Single-PSF-lifetime} L. Radzihovsky, P. B. Weichman, and J. I. Park, Ann. Phys. \textbf{323}, 2376 (2008).
\bibitem{Single-PSF3}A. J. Daley, J. M. Taylor, S. Diehl, M. Baranov, and P. Zoller, Phys. Rev. Lett. \textbf{102}, 040402 (2009).
\bibitem{Single-PSF4}S. Diehl, M. Baranov, A. J. Daley, and P. Zoller, Phys. Rev. Lett. \textbf{104}, 165301 (2010).
\bibitem{pair_hopping1}Y. C. Wang, W. Z. Zhang, H. Shao and W. A. Guo,  Chin. Phys. B \textbf{22}, 96702 (2013).
\bibitem{pair_hopping2}W. Z. Zhang,  R. X. Yin, and Y. C. Wang, Phys. Rev. B \textbf{88}, 174515 (2013).
\bibitem{pair_hopping3}Q. Z. Zhu, Q. Zhang, and B. Wu, J. Phys. B: At. Mol. Opt. Phys. \textbf{48}, 045301 (2015).
\bibitem{pair_hopping4}O. J\"{u}rgensen, K. Sengstock, and D. S. L\"{u}hmann, Sci. Rep. \textbf{5}, 12912 (2015).
\bibitem{non_standard}O. Dutta, M. Gajda, P. Hauke, M. Lewenstein, D. S. L\"{u}hmann, B. A. Malomed, T. Sowi\'{n}ski, and
J. Zakrzewski, Rep. Prog. Phys. \textbf{78}, 066001 (2015).


\bibitem{floquet2}A. Eckardt, and E. Anisimovas, New. J. Phys. \textbf{17}, 093039 (2015).
\bibitem{Bukov} M. Bukov, L. D¡¯Alessio, and A. Polkovnikov, Adv. Phys. \textbf{64}, 139 (2015).
\bibitem{Goldman}N. Goldman, and J. Dalibard, Phys. Rev. X \textbf{4}, 031027 (2014).
\bibitem{mei}F. Mei, Jia-Bin You, Dan-Wei Zhang, X.C. Yang, R. Fazio, Shi-Liang Zhu, L. C. Kwek, Phys. Rev. A \textbf{90}, 063638 (2014).
\bibitem{tunable-so}K. Jim\'{e}nez-Garc\'{\i}a, L. J. LeBlanc, R. A. Williams, M. C. Beeler, C. Qu, M. Gong, C. Zhang, and I. B. Spielman,
 Phys. Rev.Lett.\textbf{ 114}, 125301 (2015).
\bibitem{induce} X. Luo, L. Wu, J. Chen, Q. Guan, K. Gao, Z. F. Xu, L. You, and R. Wang, Sci. Rep. \textbf{6}, 18983 (2016).

\bibitem{s-hopping1} T. Keilmann, S. Lanzmich, I. McCulloch, and m. Roncaglia, Nat. Commun. \textbf{2}, 361 (2011).
\bibitem{s-hopping2} \'{A}. Rapp, X. L. Deng, and L. Santos, Phys. Rev. Lett. \textbf{109}, 203005 (2012).
\bibitem{s-hopping3}S. Greschner, L. Santos, and D. Poletti, Phys. Rev. Lett. \textbf{113}, 183002 (2014).
\bibitem{s-hopping4} S. Greschner, G. Sun, D. Poletti, and L. Santos Phys. Rev. Lett. \textbf{113}, 215303 (2014).
\bibitem{s-hopping5} S. Greschner and L. Santos, Phys. Rev. Lett. \textbf{115}, 053002 (2015).
\bibitem{s-hopping6} C. Str\"{a}ter, S. C. L. Srivastava, and A. Eckardt, Phys. Rev. Lett. \textbf{117}, 205303 (2016).
\bibitem{s-hopping7} F. Meinert, M. J. Mark, K. Lauber, A. J. Daley, and H. C. N\"{a}gerl, Phys. Rev. Lett. \textbf{116}, 205301 (2016).
\bibitem{s-hopping8} L. W. Clark, B. M. Anderson, L. Feng, A. Gaj, K. Levin, and C. Chin,  Phys. Rev. Lett. \textbf{121}, 030402 (2018).
\bibitem{s-hopping9}M. Messer, K. Sandholzer, F. G\"{o}rg, J. Minguzzi, R. Desbuquois, and T. Esslinger, Phys. Rev. Lett. \textbf{121}, 233603 (2018).
\bibitem{s-hopping10}F. G\"{o}rg, M. Messer, K. Sandholzer, G. Jotzu, R. Desbuquois, and T. Esslinger, Nature(London) \textbf{553}, 481 (2018).
\bibitem{s-hopping11}L. Barbiero, C. Schweizer, M. Aidelsburger, E. Demler, N. Goldman, F. Grusdt, Sci. Adv. \textbf{5}, eaav7444 (2019).
\bibitem{s-hopping12} F. G\"{o}rg, K. Sandholzer, J. Minguzzi, R. Desbuquois, M. Messer, and
T. Esslinger, Nature Physics, \textbf{15}, 1161 (2019).
\bibitem{s-hopping13} C. Schweizer, F. Grusdt, M. Berngruber, L. Barbiero, E. Demler, N. Goldman, I. Bloch and M. Aidelsburger, Nature Physics, \textbf{15}, 1168 (2019).


\bibitem{Synthetic_D}A. Celi, P. Massignan, J. Ruseckas, N. Goldman, I. B. Spielman, G. Juzeli\={u}nas, and M. Lewenstein, Phys. Rev. Lett. \textbf{112}, 043001 (2014).

\bibitem{rf}J. Struck, J. Simonet, and K. Sengstock, Phys. Rev. A \textbf{90}, 031601(R) (2014).
\bibitem{C-sf-mi1} A. Eckardt, C. Weiss, and M. Holthaus, Phys. Rev. Lett. \textbf{95}, 260404 (2005).
\bibitem{supplemental}See the supplemental materials for more details on the derivation of the effective Hamiltonian, and the possible phases.
\bibitem{Gutzwiller1}D. Jaksch, C. Bruder, J. I. Cirac, C. W. Gardiner, and P. Zoller, Phys. Rev. Lett. \textbf{81}, 3108 (1998).
\bibitem{Gutzwiller2}W. Zwerger, J. Opt. B \textbf{5}, S9 (2003).
\bibitem{Gutzwiller3}C. Trefzger, C. Menotti, B. Capogrosso-Sansone, and M. Lewenstein, J. Phys. B \textbf{44}, 193001 (2011).
\bibitem{Gutzwiller4}D. Jaksch, V. Venturi, J. I. Cirac, C. J. Williams, and P. Zoller, Phys. Rev. Lett. \textbf{89}, 040402 (2002).
\bibitem{Gutzwiller5}U. Bissbort, S. G\"{o}tze, Y. Li, J. Heinze, J. S. Krauser, M. Weinberg, C. Becker, K. Sengstock, and W. Hofstetter, Phys. Rev. Lett. \textbf{106}, 205303 (2011).
\bibitem{Gutzwiller6} U. R. Fischer and B. Xiong, Phys. Rev. A \textbf{84}, 063635 (2011).
\bibitem{excitation} K. V. Krutitsky, and P. Navez,  Phys. Rev. A \textbf{84}, 033602 (2011).
\bibitem{Gutzwiller7}Dirk-S\"{o}ren L\"{u}hmann, Phys. Rev. A \textbf{87}, 043619 (2013).
\bibitem{even_coherent} V. V. Dodonov, I. A. Malkin, and V. I. Man'ko, Physica \textbf{72}, 597 (1974)
\bibitem{odd_coherent} M. Hillery, Phys. Rev. A \textbf{36}, 3796 (1987).
\bibitem{spin-removal}M. F. Parsons, A. Mazurenko, C. S. Chiu, G. Ji, D. Greif and M. Greiner, Science  \textbf{353}, 1253 (2016).
\bibitem{spin-removal1} P. T. Brown, D. Mitra, E. Guardado-Sanchez, P. Schau{\ss},
S. S. Kondov, E. Khatami, T. Paiva, N. Trivedi, D. A. Huse, W. S. Bakr, Science  \textbf{357}, 1385 (2017).
\bibitem{In-situ-imaging}H. Ott,  Rep. Prog. Phys. \textbf{79}, 054401 (2016).
\bibitem{In-situ-B-MI1}N. Gemelke, X. Zhang, C. Hung, and C. Chin, Nature (London) \textbf{460}, 995 (2009).
\bibitem{In-situ-B-MI2}W. S. Bakr, A. Peng, M. E. Tai, R. Ma, J. Simon, J. I. Gillen, S. F\"{o}lling, L. Pollet, M. Greiner,  Science  \textbf{329}, 547 (2010).
\bibitem{In-situ-B-MI3}J. F. Sherson, C. Weitenberg, M. Endres, M. Cheneau, I. Bloch and S. Kuhr, Nature (London) \textbf{467}, 68 (2010).
\bibitem{In-situ-F-MI1}D. Greif, M. F. Parsons, A. Mazurenko, C. S. Chiu, S. Blatt, F. Huber, G. Ji, M. Greiner, Science
 \textbf{351}, 953 (2016).
 \bibitem{In-situ-F-MI2}L. W. Cheuk, M. A. Nichols, K. R. Lawrence, M. Okan, H. Zhang, and M. W. Zwierlein, Phys. Rev. Lett. \textbf{116}, 235301 (2016).
 \bibitem{spin-resolved-TOF}Z. Wu, L. Zhang, W. Sun, X. T. Xu, B. Z. Wang, S. C. Ji, Y. J. Deng,  S. Chen, X. J. Liu, J. W. Pan, Science \textbf{354}, 83 (2016).
\end{thebibliography}
\end{document}